\documentstyle[aps,pre,psfig]{revtex}
\begin{document}

\title{Nonmonotonic roughness evolution in unstable growth}
\author{Claudio Castellano and Joachim Krug}
\address{
Fachbereich Physik, Universit\"at GH Essen, 45117 Essen, Germany \\
}
\date{\today}
\maketitle

\begin{abstract}
The roughness of vapor-deposited thin films can display a nonmonotonic
dependence on film thickness, if the smoothening of the small-scale
features of the substrate dominates over growth-induced 
roughening in the early stage of evolution. We present a detailed
analysis of this phenomenon in the framework of the continuum theory
of unstable homoepitaxy. Using the spherical approximation of 
phase ordering kinetics, the effect of nonlinearities and noise can be
treated explicitly. The substrate roughness is characterized by
the dimensionless parameter $Q = W_0/(k_0 a^2)$,
where $W_0$ denotes the roughness amplitude, $k_0$ is the small scale
cutoff wavenumber of the roughness spectrum, and $a$ is the lattice 
constant. Depending on $Q$, the diffusion length $l_D$ and the
Ehrlich-Schwoebel length $l_{ES}$, five regimes are identified
in which the position of the roughness minimum is determined by
different physical mechanisms. The analytic estimates are compared
by numerical simulations of the full nonlinear evolution equation. 

\end{abstract}


\section{Introduction}

The morphology of thin film surfaces has a decisive influence
on many film properties. The control of growth-induced surface
roughness is therefore a central concern in thin film science
and technology. Two types of roughening mechanisms have been
extensively studied in recent years \cite{Barabasi,Krug97}. The term
{\em kinetic roughening} is commonly used to refer 
to a {\em stochastic} mechanism, 
in which fluctuations in the deposition flux interact with
thermal smoothening to generate a scale-invariant, rough
morphology. This theoretically appealing but empirically rather
elusive phenomenon \cite{Krim95} is often superseded by a second, 
{\em deterministic} mechanism, a 
{\em growth instability} associated with reduced interlayer
transport and slope-dependent mass currents along the surface
\cite{Krug97,Villain91,Politi99}. The hallmark of unstable growth is
a morphology of more or less regular mounds with a clearly
developed characteristic length scale. While in practice
the distinction between the two types of roughening mechanisms
may not always be so clear-cut \cite{Siegert96}, 
they are very different conceptually. 

In addition to the growth-induced roughness, clearly also 
the roughness of the substrate affects the film morphology.
Since the growing film covers up the small-scale details of
the substrate modulations, the substrate contribution to the
roughness is expected to decrease with increasing film
thickness, while the growth-induced roughness component increases.
Under suitable conditions this leads to the somewhat counterintuitive
possibility of a {\em minimum} of the total surface roughness 
at a nonzero film thickness. This phenomenon has been observed
in several growth experiments \cite{Koenig95,Klemradt96,Gyure98},
and a theoretical description 
has been worked out on the level of linear continuum theories
of kinetic roughening \cite{Majaniemi96} and unstable growth
\cite{Krug99}. 

In Ref.\cite{Krug99} a quantitative comparison with
the experiments of Gyure et al.\cite{Gyure98} was attempted,
which indicated an important influence of nonlinearities. 
This motivated the present study, in which the nonlinear term
in the growth equation is treated explicitly using the spherical
approximation of phase ordering kinetics \cite{Bray94,Rost97}. 
We find that the interplay of instability, nonlinearity and noise
gives rise to a rather complex behavior, in which the position of 
the roughness minimum can be determined by several distinct 
physical mechanisms. For a quick overview of the different
regimes we refer the curious reader to Table I.

The paper is organized as follows. In the next section we introduce
the standard continuum equation for unstable homoepitaxial growth
\cite{Rost97,Krug99a} and describe the strategy for its 
analytical solution. Section \ref{Det} is devoted to the roughness
evolution in the absence of noise. We first recapitulate the
linear analysis of Ref.\cite{Krug99}, then provide a detailed
analysis of the relevance of the nonlinearity and the nonlinear
behavior, and finally discuss the influence of correlated initial
roughness. The effects of noise are analyzed in Section \ref{Noise}.
In Section \ref{Numerics} we compare the analytic estimates to 
a numerical evaluation of the spherical approximation, as well as
to numerical simulations of the full nonlinear growth equation, finding
good agreement in all cases. Finally, some conclusions are formulated
in Section \ref{Discussion}.

\section{The continuum equation}
The evolution of a surface growing under typical Molecular Beam
Epitaxy (MBE) conditions
is described by an equation of the form \cite{Villain91}
\begin{equation}
\partial_t H + \nabla \cdot {\bf J} = a_\perp F + \eta,
\label{eq1}
\end{equation}
where $H$ is the height, ${\bf J}$ is the surface current,
$a_\perp$ is the monolayer thickness, $F$ is
the average value of the deposition flux and $\eta$ is a noise term
describing fluctuations in the flux (shot noise) and in the diffusion
of adatoms on the surface.
The constant term $a_\perp F$ can be eliminated passing to the 
frame of reference $H= a_\perp Ft +h$ moving with the average height.
The noise has zero average and correlations 
\begin{equation}
\langle \eta({\bf x},t)
\eta({\bf x'},t') \rangle
= (2\pi)^{-2} \delta(t-t') (R_S-R_D \nabla^2) \delta({\bf x}-{\bf x'}),
\end{equation}
with the amplitudes $R_S$ and $R_D$ 
representing the effect of shot and diffusion noise,
respectively.

The surface current is the sum of two contributions
\begin{equation}
{\bf J} = \kappa \nabla (\nabla^2 h) + f[(\nabla h)^2] \nabla h.
\label{J}
\end{equation}
The first tends to smoothen the surface and has the form of a
capillarity term~\cite{Mullins}, even though it may be dominated
by nonequilibrium effects such as nucleation \cite{Krug99a,Politi96}.
The second term models a growth-induced surface current, whose
existence is often (but not
necessarily~\cite{PierreLouis99,Ramana99,Politi00}) caused by the presence
of an Ehrlich-Schwoebel barrier for interlayer diffusion.
We assume in-plane isotropy of the current (for a discussion of origins
and consequences of anisotropy see \cite{Politi00,Siegert98}). 
For the function $f(a)$ we use the form valid in the weak barrier
limit~\cite{Politi99}
\begin{equation}
\label{f}
f(a) =
{\alpha \over ( 1 + l_D \sqrt{a}/a_\perp) (1+l_{ES} \sqrt{a}/a_\perp)},
\label{current}
\end{equation}
where $l_D$ is the diffusion length and $l_{ES}$ the Ehrlich-Schwoebel
length. These length scales are related to the in-layer hopping
rate $D$, the inter-layer hopping rate $D'$ and the deposition flux
$F$ through 
\begin{equation}
\label{lengths}
l_D \approx (D/F)^\gamma a_\parallel, \;\;\;\; 
l_{ES} = (D/D' - 1) a_\parallel.
\end{equation}
Here $a_\parallel$ denotes the in-layer lattice constant and the
exponent $\gamma$ depends on the size of the critical cluster
for two-dimensional nucleation \cite{Villain92}. Conditions of
weak and strong step edge barriers can be distinguished according
to whether $l_{ES} \gg l_D$ (strong barriers) or 
$l_{ES} \ll l_D$ (weak barriers) \cite{Krug97,Politi99,Politi96}.
Here we focus on the latter case, in which a continuum description
is most likely to be valid.
 
The coefficients $\alpha$ and $\kappa$ 
in Eqs.(\ref{J},\ref{f}) are related to microscopic parameters
by $\alpha \approx F l_{ES} l_D/2$, $\kappa \approx F l_D^4$
\cite{Krug97,Politi99,Krug99a,Politi96}. 
We will for simplicity
assume that  the equality sign holds in these formulas; however numerical
factors are not precisely known and the equalities should be intended instead
only in order of magnitude.
Also the amplitude of the noise terms is connected to microscopic
parameters through $R_D = l_D^2 R_S = l_D^2 F a_\perp^2 a_\parallel^2$.

By inserting the expression 
(\ref{J}) of the current in the equation for the height
and neglecting noise one obtains
\begin{equation}
\partial_t h = - \kappa (\nabla^2)^2 h - \nabla [f((\nabla h)^2) \nabla h].
\label{eqh}
\end{equation}
This strongly nonlinear equation is reminiscent of the Cahn-Hilliard
equation for phase-ordering in systems with conserved order parameter
\cite{Bray94}.
A widely used method for investigating this kind of nonlinear evolution
is the large-$N$ limit, or spherical approximation.
In the present context it consists in replacing the argument of the
nonlinear current $f$ with its average value
$a(t)=\langle (\nabla h)^2 \rangle$.
In this way, Eq.~(\ref{eqh}) is effectively linearized.
It is then possible to write down a closed form linear equation for
the structure factor
$S(k,t) = \langle {\hat h}({\bf k},t) {\hat h}(-{\bf k},t)\rangle$

\begin{equation}
\partial_t S(k,t) = -2 \left[\kappa k^4 - f[a(t)] k^2 \right] S(k,t),
\label{eqS}
\end{equation}
where $a(t)$ must be determined self-consistently

\begin{equation}
a(t) =  2\pi \! \int_0^{k_0} \! \! dk \; k^3 S(k,t).
\label{a}
\end{equation}

The solution of this pair of coupled equations has already been derived
for long times by Rost and Krug~\cite{Rost97}.
Here we concentrate on the short time behavior, i.e. all what happens
before the instability sets in. In this time range we expect the
large-$N$ approximation to give a fairly accurate description of the
nonlinear behavior~\cite{Castellano98}, since correlations are still
small in range and amplitude.
In particular we will be interested in
the time evolution of the surface roughness
\begin{equation}
W^2(t) =  2\pi \! \int_0^{k_0} \! \! dk \; k S(k,t).
\end{equation}

We usually 
assume as initial condition $S_0(k) =S(k=0,t)$ a white spectrum with an
upper cutoff $k_0 =\pi /l_0$
\begin{equation}
\label{S0}
S_0(k) = \left\{ \begin{array}{l@{\quad for \quad}l}
W_0^2/(\pi k_0^2) & k < k_0 \\
0 & k > k_0, \end{array} \right.
\end{equation}
which implies $W^2(0)=W_0^2$ and $a(0) = W_0^2 k_0^2 /2$.
The dimensionless number
\begin{equation}
\label{Q}
Q = W_0/(k_0 a_\perp a_\parallel)
\end{equation}
will turn out to provide a useful measure for the strength of the
initial roughness; note that it involves both the amplitude
($W_0$) and the small scale cutoff. Other types of initial roughness
spectra will be treated in Section \ref{Corr}.

\section{Solution in the deterministic case}
\label{Det}

Equation~(\ref{eqS}) can be formally integrated
\begin{equation}
\label{formal}
S_{det}(k,t) = 2 \left({W_0 \over k_0}\right)^2
\exp \left[ -2 \kappa k^4 t + 2 k^2 b(t) \right],
\end{equation}
where
\begin{equation}
b(t) \! = \! \int_0^t ds \; f[a(s)].
\label{b}
\end{equation}

By defining $k_m^2(t) = b(t)/(2 \kappa t)$ one can rewrite
\begin{eqnarray}
W^2(t)& = &  2 \left( W_0 \over k_0 \right)^2 \int_0^{k_0} \! \! dk \; k
\exp \left\{2t \kappa k^4\left[2\left({k_m \over k}\right)^2-1\right] 
\right\} \\
& =&
W_0^2 \sqrt{\pi \over 2} \exp(2 \kappa k_m^4 t){[{\rm Erf}(k_m^2
 \sqrt{2 \kappa t}) + {\rm Erf}((k_0^2-k_m^2) \sqrt{2 \kappa t})]
\over 2 \sqrt{\kappa t} k_0^2}
\label{W2alltimes}
\end{eqnarray}
with ${\rm Erf}(s) = (2/\sqrt{\pi}) \int_0^s \exp(-t^2) dt$. 
The wavenumber 
$k_m$ is the position of the structure function peak (when it is a real
number, otherwise the peak is for $k=0$).
Within linear theory its value is constant, $k_m=k_l \equiv
\sqrt{\alpha/2\kappa}$,
while in general $k_m$ is a function of time.

The derivative of $W^2$ with respect to $t$ at $t=0$ is
\begin{equation}
\left. {dW^2(t) \over dt} \right|_{t=0} =
{2 \over 3} W_0^2 \kappa k_0^2 (3 {\tilde k}^2-k_0^2),
\end{equation}
where ${\tilde k} \equiv k_m(0)=\sqrt{f[a(0)]/2 \kappa} =
\sqrt{f(W_0^2 k_0^2/2)/2 \kappa}$.
Hence if ${\tilde k}/k_0 > 1/\sqrt{3}$ the roughness grows from the
beginning and there is no minimum in the behavior of $W^2(t)$:
The instability is immediately at work.
Notice however that
the condition for the existence of the minimum involves ${\tilde k}$ and
not $k_l$.
When the initial roughness is large ${\tilde k} \ll k_l$ and it may
occur that ${\tilde k} \ll k_0 \ll k_l$: In such a case a minimum occurs even
if the linear theory does not predict it.
This fact, together with the observation that $k_l/k_0$ does not depend on
$W_0$, implies that if $k_0$ is sufficiently large a minimum exists
even for very small amplitude of the initial fluctuations.
If $k_0$ is not large, only a strong initial roughness can originate
a non monotonic behavior of the width.

We will assume in the following that a minimum exists.
To study the detailed behavior of the system one should in principle
consider Eq.~(\ref{formal}) and~(\ref{b}) simultaneously.
However, expanding the expression of $W^2(t)$ for small $k_m/k_0$
\begin{equation}
W^2(t) = \left( {W_0 \over k_0} \right)^2
\sqrt{\pi \over 2} {{\rm Erf}(k_0^2 \sqrt{2\kappa t})
\over 2 \sqrt{\kappa t}} + W_0^2 [1-\exp(-2\kappa k_0^4 t)]
\left({k_m \over k_0}
\right)^2 + O\left({k_m \over k_0} \right)^4,
\label{Wexpans}
\end{equation}
one finds that after a transient time $t_0 = 1/(2 \kappa k_0^4)$ the width
starts decreasing as $1/\sqrt{\kappa t}$ and it does so until
$1/\sqrt{\kappa t} \simeq k_m^2$.
During this time interval the width decreases in time and depends
only on $\kappa$ and $k_0$, not on $k_m$ (hence not on the form of the
current, not even its linear expansion).
The system is effectively described by an evolution equation~(\ref{eqh})
where only the relaxational term proportional to $\kappa$ 
matters~\cite{Majaniemi96,Corberi95}.
This fact is crucial for all the following calculations: The structure
factor is known
\begin{equation}
S_{det}(k,t) = S_0 \exp(-2 \kappa k^4 t)
\label{Sdecay}
\end{equation}
and can be used to compute $a(t)$ and $k_m(t)$, which change nonlinearly
in time, but do not affect significantly $S_{det}(t)$.
This situation persists up to a time such that other terms in the
expansion~(\ref{Wexpans}) become large.
Since the other terms grow with $t$, it is around this time that
$W^2(t)$ reaches a minimum.
The only role played by the current $f$ is to determine the time evolution
of $k_m(t)$ and hence when the decay of the initial fluctuations ends:
But $f$ does not affect the way $W^2(t)$ decreases.

The form~(\ref{Sdecay}) of the structure factor may be interpreted as that
of a system which is coarsening with a typical correlation length growing
as $L(t) \sim (\kappa t)^{1/4}$. 
The initial condition creates ``domains'' of size $k_0^{-1}$, much smaller
than the length of the instability $k_m^{-1}$: the system evolves by reducing
the amplitude of fluctuations and increasing the correlation length.
This explains why the initial decrease of $W^2$ is seen only for small
${\tilde k}/k_0$.
This coarsening process continues until
\begin{equation}
L \simeq k_m^{-1}.
\label{mintime}
\end{equation}
>From this time on the evolution proceeds by amplifying
fluctuations of scale $L$ and the instability sets in.
Notice that while in the linear case $k_m=k_l$ is constant in time,
when nonlinearities are taken into account $k_m$ grows in time but
remains always smaller than $k_l$,
because $f(a) \leq \alpha$ for all $a$. 
Therefore the time where the minimum
occurs in the linear theory is a lower bound for the same quantity
in the nonlinear case.

We now compute in detail how the position of the minimum depends on the
amplitude of the initial fluctuations.
For reference it is useful to summarize first the results of the
linear theory~\cite{Krug99}.

\begin{subsection}{Linear theory}

The assumption of linearity for the current implies $a(t)=0$.
Hence $f[a(t)]=f[0] =
\alpha$, $b(t)=\alpha t$ and $k_m^2 = k_l^2 = \alpha /2\kappa$.
Then the temporal evolution of the width is fully specified for all times by
Eq.~(\ref{W2alltimes}).
Letting $k_0 \to \infty$ such a formula can be cast as
\begin{equation}
W^2(t) = {W_0^2 \over 4} \left(k_l \over k_0 \right)^2 \Phi(t/\tau_l),
\label{scaling}
\end{equation}
where 
\begin{equation}
\label{taulinear}
\tau_l = 4 \kappa/\alpha^2 
\end{equation} 
is the inverse amplification rate of
the maximally unstable fluctuations and the scaling function
is
\begin{equation}
\label{Phi}
\Phi(x) = e^{2x}\sqrt{2\pi/x}[1 + {\rm Erf}(\sqrt{2x})].
\end{equation}
This formula is valid only for times greater than
$t_0 = 1/2\kappa k_0^4 \sim (k_l/k_0)^4 \tau_l \ll \tau_l$.

The width attains a minimum at a time
\begin{equation}
t_{\rm min}^l \approx 0.18\; \tau_l,
\label{tmin1}
\end{equation}
where it has been reduced by a factor 
\begin{equation}
\label{Wmin}
W^2(t_{min}^l)/W_0^2 \approx 3.42 \; (k_l/k_0)^2.
\end{equation}
This minimum marks the transition between the initial power-law decrease
and the eventual exponential increase due to the linear instability.
\end{subsection}

\begin{subsection}{Nonlinear theory}

The initial condition implies that $a(0)=W_0^2 k_0^2 /2$.
Then, for very small $t$
\begin{equation}
b(t) = f(W_0^2 k_0^2/2) t \equiv {\tilde \alpha} t.
\label{btilde}
\end{equation}
By comparing ${\tilde \alpha}$ with $\alpha$ one recovers the condition
(21) of Ref.~\cite{Krug99} for the irrelevance of the nonlinearity
\begin{equation}
{W_0 k_0 l_D \over a_\perp} \ll 1.
\label{21Krug}
\end{equation}
However, this condition turns out to be too restrictive.
The reason is that it assesses the relevance of the nonlinearity from
its importance in the expression of the unstable current at the initial
time. But, as discussed above, $f$ does not play any role in the
initial evolution of the width.
The relevance of the nonlinearity must instead be established from its
influence on the position of the minimum of $W^2(t)$, i.e. for long times.
By that time the initial fluctuations have already been reduced
significantly.

Expanding $a(t)$ as a function of $k_m/k_0$
\begin{eqnarray}
a(t) & = & {W_0^2 \over 4 \kappa k_0^2 t} [1-\exp(-2\kappa k_0^4 t)] \\
 & + & {W_0^2 \over 4} \left[-4 \exp(-2\kappa k_0^4 t) k_0^2 +
\sqrt{2\pi \over \kappa t}
{\rm Erf}(k_0^2 \sqrt{2\kappa t}) \right] \left( k_m \over k_0 \right)^2
+ O\left( k_m \over k_0\right)^4.
\end{eqnarray}
one can see that $a(t)$ is constant only for times of the order of
$t_0 =1/(2 \kappa k_0^4)$, indicating that Eq.~(\ref{btilde}) soon
loses validity.

For longer times (up to $t \simeq 1/[\kappa k_m(t)^4]$), 
\begin{equation}
a(t) = {W_0^2 \over 4 \kappa k_0^2 t}.
\label{a2}
\end{equation}
Inserting Eq.~(\ref{a2}) in the expression~(\ref{b}) we can compute the
long time behavior of $b(t)$
\begin{eqnarray}
b(t) & = & \int_0^t ds \;
{\alpha \over ( 1 + {l_D W_0 \over 2 a_\perp k_0 \sqrt{\kappa s}})
( 1 + {l_{ES} W_0 \over 2 a_\perp k_0 \sqrt{\kappa s}})} \\
& = &\! \int_0^t ds \;
 {\alpha \over (1+ \sqrt{\tau_{D} /s})(1+\sqrt{\tau_{ES}/s})},
\label{bexplicit}
\end{eqnarray}
where we have introduced two timescales
\begin{equation}
\label{tauD}
\tau_D = \left({l_D W_0 k_l^2 \over 2 a_\perp k_0} \right)^2 \tau_l
= Q^2 y^2 \left({a_\parallel \over 8 l_D}\right)^2 \tau_l
\end{equation}
and
\begin{equation}
\label{tauES}
\tau_{ES} = \left({l_{ES} W_0 k_l^2 \over 2 a_\perp k_0} \right)^2 \tau_l
= Q^2 y^4 \left({a_\parallel \over 8 l_D}\right)^2 \tau_l.
\end{equation}
In the right equalities we have introduced the quantity 
$y=l_{ES}/l_D$ and $Q$ is defined in (\ref{Q}).
It is important to stress that since we are considering weak Ehrlich-Schwoebel
barriers $y \ll 1$ and hence $\tau_{ES} \ll \tau_{D}$.
Moreover, realistic values of the diffusion length are such that
$l_D / a_\parallel \gg 1$.
Depending on the initial roughness via the parameter $Q$, the timescales
$\tau_{ES}$ and $\tau_D$ will be larger or smaller than the initial one,
$\tau_l$, giving rise to different scenarios.

\begin{subsubsection}{Irrelevant nonlinearity (intermediate initial roughness)}

Consider first the case of a fairly small initial roughness,
i. e. $Q y a_\parallel /8 l_D \ll 1$, so that 
$\tau_{ES} \ll \tau_D \ll \tau_l$.
Then for $t_0 \ll t \ll \tau_{ES}$ one can neglect the constant
term in the denominator of Eq.~(\ref{bexplicit})
\begin{equation}
b(t) \approx  \! \int_0^t ds \; {\alpha s \over \sqrt{\tau_D \tau_{ES}}}=
{\alpha t^2 \over 2 \sqrt{\tau_D \tau_{ES}}}.
\label{bES}
\end{equation}
For $\tau_{ES} \ll t \ll \tau_{D}$ instead
\begin{equation}
b(t) \approx \! \int_0^t ds \; {\alpha  s^{1/2} \over \sqrt{\tau_D}} =
{2 \alpha t^{3/2} \over 3 \sqrt{\tau_D}},
\label{bD}
\end{equation}
while for $\tau_{D} \ll t \ll \tau_{l}$
\begin{equation}
b(t) \approx \alpha t.
\end{equation}

$b(t)$ undergoes several changes during the time
evolution passing through two intermediate behaviors.
However it is easy to see that these variations in the form of $b(t)$
are too short lived to affect the time evolution of $W^2$ (or $a(t)$).
The minimum width is reached when the
effect of the nonlinearity is already lost, and is well described by
linear theory. 
The condition for the irrelevance of the nonlinearity is therefore
that $\tau_D \ll \tau_l$, that is
\begin{equation}
{W_0 k_0 l_D \over 2 a_\perp} \left({k_l \over k_0}\right)^2 \ll 1.
\label{21nonlin}
\end{equation}
Comparing with Eq.~(\ref{21Krug}) it is clear that the relevance 
of the nonlinearity is strongly reduced when $k_l \ll k_0$.
Notice moreover that the role of $k_0$ is opposite compared to the
condition~(\ref{21Krug}): An initial substrate rough down to very small
length scales (large $k_0$) makes the nonlinearity less relevant.

\end{subsubsection}

\begin{subsubsection}{Relevant nonlinearity (large initial roughness)}

Let us assume instead $\tau_{ES} \ll \tau_l \ll \tau_D$, i.e.
$8 l_D/(y a_\parallel) \ll Q \ll 8 l_D/(y^2 a_\parallel)$.
One has for $\tau_{ES} \ll t \ll \tau_D$
\begin{equation}
b(t)= {2 \alpha \over 3 \sqrt{\tau_D}} t^{3/2} \ll \alpha t.
\label{bB}
\end{equation}
With this expression of $b(t)$ the value of $k_m$ is much smaller
than the linear value $k_l$ and the condition~(\ref{mintime}) for the
minimum is attained on time scales larger than $\tau_l$.
To estimate $t_{min}$ more precisely 
the form of $W^2(t)$ can be written in the scaling form 
\begin{equation}
W^2(t) = {W_0^2 \over 4} \left(k_m(t) \over k_0 \right)^2 \Phi(t \kappa
k_m(t)^4),
\end{equation}
where $\Phi$ is defined in (\ref{Phi}).
Now the minimum value of $W^2$ is not reached where $\Phi'(x)=0$, because
$k_m$ depends on $t$.
The condition for the minimum of $W^2$ is instead
\begin{equation}
x \Phi'(x) (2 b't-b) + \Phi(x) (b't -b)=0
\end{equation}
Using the expression~(\ref{bB}) for $b(t)$ one gets
\begin{equation}
{x \Phi'(x) \over \Phi(x)}=-{1 \over 4}
\end{equation}
whose solution is $x_{min} = t_{min}^D / \tau(t_{min}^D) \approx 0.06$
yielding
\begin{equation}
t_{min}^D \approx 0.4 (\tau_l \tau_D)^{1/2} =
0.4 \; Q y \left({a_\parallel \over 8 l_D} \right) \tau_l
\label{tmin2}
\end{equation} 
Hence $t_{min}^D$ is larger than $\tau_l$ but smaller than $\tau_D$.
The value of the width at the minimum is
\begin{equation}
W(t_{min}^D)/W_0 \approx 1.3 \left({k_l \over k_0}\right)
\left({\tau_{l} \over \tau_D}\right)^{1/2}
\end{equation}

If instead $Q \gg 8 l_D/(y^2 a_\parallel)$, then
$\tau_l \ll \tau_{ES} \ll \tau_D$. By means of
an analogous procedure, one finds that the minimum occurs for
$x_{min} \approx 0.03$
yielding
\begin{equation}
t_{min}^{ES} \approx 0.5 \left(\tau_l \tau_D \tau_{ES}\right)^{1/3} =
0.5 \; Q^{4/3} y^2 \left({a_\parallel \over 8 l_D}\right)^{4/3} \tau_l
\label{tmin3}
\end{equation}
The value of the width at the minimum is
\begin{equation}
W(t_{min}^{ES})/W_0 \approx 1.1 \left({k_l \over k_0}\right)
\left({\tau_{l}^2 \over \tau_D \tau_{ES}}\right)^{1/3}.
\end{equation}

\end{subsubsection}
\end{subsection}

\subsection{Correlated initial conditions}
\label{Corr}

The previous results can be easily extended to the case of a substrate
with correlated roughness, i.e. with
\begin{equation}
S_0(k) = \left\{ \begin{array}{l@{\quad for \quad}l}
A & k < k^* \\
A \left(k^*/k\right)^{\theta} & k^* < k < k_0 \\
0 & k > k_0, \end{array} \right.
\end{equation}
with $\theta>0$, and $k_0/k^* \gg 1$.

In this case the condition for an initial decrease of the roughness is,
in the limit $k_0/k^* \to \infty$,
\begin{equation}
\left({{\tilde k} \over k_0} \right)^2 < 
\left\{ \begin{array}{l@{\quad for \quad}l}
{4-\theta \over 2(6-\theta)} & \theta < 4 \\
0 & \theta > 4 \end{array} \right.
\end{equation}

Hence for $\theta > 4$ the width can only increase monotonically from the
beginning.

\subsubsection{Linear theory}
For $\theta<2$ one can safely take $k^* \to 0$ and find, for $t \gg
1/(2 \kappa k_0^4)$ and to second order in $k_l$,
\begin{equation}
W^2(t) = \pi A k^{*\theta} (2 \kappa t)^{\theta/4} \left[ 
{1 \over (8 \kappa t)^{1/2}} \Gamma\left({2 - \theta \over 4}\right)
+ k_l^2 \Gamma\left(1 - {\theta \over 4}\right)
\right].
\label{W2corr}
\end{equation}
Estimating the position of the minimum from the time when the second
term equals the first, one obtains
\begin{equation}
t_{min}^l(\theta) = {1 \over 8 \kappa k_l^4} \left[
{\Gamma\left({2-\theta \over 4}\right) \over \Gamma \left(1-{\theta \over
4} \right)}
 \right]^2,
\end{equation}
while the minimum width reached is
\begin{equation}
{W^2(t_{min}^l(\theta)) \over W_0^2} = (2-\theta) 2^{-(\theta/2 + 1)}
\Gamma\left({2-\theta \over 4} \right)^{\theta/2}
\Gamma\left(1-{\theta \over 4} \right)^{1-\theta/2}
\left({k_l \over k_0} \right)^{2-\theta}.
\end{equation}
Both $t_{min}^l(\theta)$ and $W^2(t_{min}^l(\theta))$
are growing functions of $\theta$:
The minimum is delayed and made shallower by the presence of correlations
in the roughness. This occurs because
the roughness is concentrated on large length scales and the damping
of small scale fluctuations provided by the relaxational dynamics
is less effective in the reduction of the surface width.
Such effect is most evident when $\theta$ approaches 2:
$t_{min}^l(\theta)$ diverges, while $W^2(t_{min}^l(\theta))/W_0^2$ goes to 1,
since for $\theta=2$ all the roughness is concentrated on the macroscopic
length scale $(k^*)^{-1}$.

For $2< \theta <4$, the origin of the minimum is different.
In this case one can take $k_0 \to \infty$, and find for
$t \ll 1/(2 \kappa k^{*4})$ and small $k_l$
\begin{equation}
W^2(t) = W_0^2 + 2 \pi A k^{*\theta/4} (2 \kappa t)^{\theta/4}
\left[ - {1 \over (\theta-2)(2 \kappa t)^{1/2}} +
{2 k_l^2 \over 4-\theta}.
\right]
\end{equation}
Differently from Eq.~(\ref{W2corr}), here $t$ has, in the first contribution
to the second term, a positive (and small) exponent but a negative prefactor.
Therefore in this case the initial decrease of the roughness is much
weaker and this is reflected by the minimum width, which is
close to $W_0$.
The time when the minimum is reached is 
\begin{equation}
t_{min}^l(\theta) = {1 \over 8 \kappa k_l^4} \left(
{4-\theta \over \theta - 2} \right)^2
\end{equation}
and vanishes, as expected, in the limit $\theta \to 4$.

\subsubsection{Nonlinear theory}

With correlated initial conditions the evolution of the average square slope,
considering $S(k,t)=S_0(k) \exp(-2 \kappa k^4 t)$ and $k^* \to 0$,
is, for $t \gg 1/(2 \kappa k_0^4)$
\begin{equation}
a(t)= {\pi A k^{* \theta} \Gamma(1-\theta/4) \over 
2 (2 \kappa t)^{1-\theta/4}}.
\end{equation}
Hence $a(t)$ is for all $\theta<4$ a decreasing function of time but its
rate of reduction vanishes as $\theta$ approaches 4.
All the previous treatment of the nonlinearity can be repeated.
The only difference in the results is that the timescales $\tau_D$ and
$\tau_{ES}$ are modified. In particular
\begin{equation}
\tau_D(\theta) = {1 \over 2K}\left[{\pi A k^{* \theta} \Gamma(1-\theta/4)
\over 2} {l_D^2 \over a_\perp^2} \right]^{4/(4-\theta)}.
\end{equation}
With this expression one can assess the relevance of the nonlinearity
by comparison with the time scale of linear theory $t_{min}^l(\theta)$.
For $\theta<2$ one finds that the nonlinearity is irrelevant
[$\tau_D(\theta) \ll t_{min}^l(\theta)$] for \\
\begin{equation}
{W_0^2 k_0^2 l_D^2 \over 4 a_\parallel^2} \left({k_l \over k_0} \right)^4
\ll \left({k_l \over k_0} \right)^\theta {1 \over (2-\theta)\Gamma(1-\theta/4)}
\left[1 -{\theta \over 2} \left({k^* \over k_0} \right)^{2-\theta} \right]
\left\{{\Gamma[(2-\theta)/4] \over 2 \Gamma(1-\theta/4)}
\right\}^{(4-\theta)/2}.
\label{21corr1}
\end{equation}

The right hand side of the previous inequality is of the order of one for
$\theta \to 0$, in agreement with Eq.~(\ref{21nonlin}).
For small $\theta$ it decreases as $(k_l/k_0)^\theta$.
For $\theta \to 2$ it diverges, as a consequence of the fact that 
$t_{min}^l(\theta)$ goes to infinity.

For $2< \theta <4$ the condition for the irrelevance of the nonlinearity
becomes
\begin{equation}
{W_0^2 k_0^2 l_D^2 \over 4 a_\parallel^2} \left({k_l \over k_0} \right)^4
\ll \left({k_l \over k_0} \right)^\theta {1 \over (2-\theta)\Gamma(1-\theta/4)}
\left[1 -{\theta \over 2} \left({k^* \over k_0} \right)^{2-\theta} \right]
\left[{4-\theta \over 2(\theta-2)} \right]^{(4-\theta)/2}.
\label{21corr2}
\end{equation}
The right hand side diverges for $\theta \to 2$ and vanishes for $\theta \to
4$, as expected since $\tau_D(\theta)$ diverges in that limit: 
For $\theta \to 4$ the nonlinearity is always relevant.

\begin{figure}
\centerline{\psfig{figure=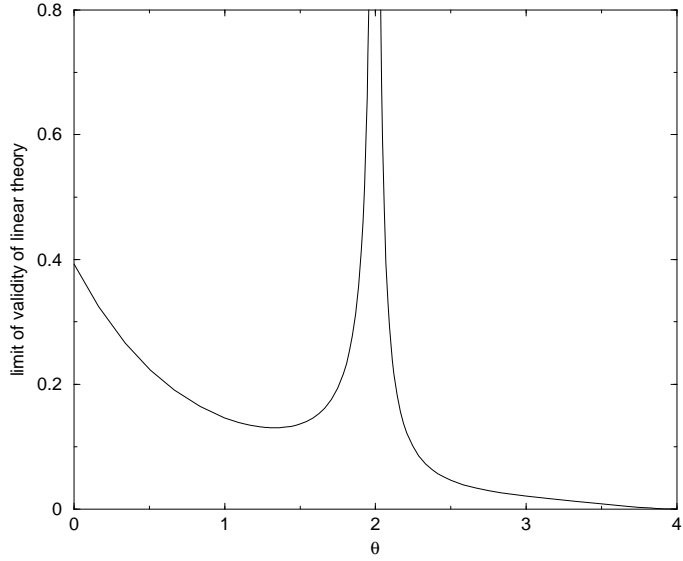,width=9cm,angle=-90}}
\caption{Plot of the right hand side of equations~(\ref{21corr1})
and~(\ref{21corr2}) vs $\theta$ for $k_l/k_0=1/10$ and $k^*/k_0=1/100$.
}
\label{Fig1}
\end{figure}

A more immediate perception of the meaning of Eqs.~(\ref{21corr1})
and~(\ref{21corr2}) is given by Fig.~\ref{Fig1}, showing the right hand
sides of the inequalities as a function of $\theta$.
Nonlinearity is irrelevant for values of
$[W_0 k_l^2 l_D / (2 a_\parallel^2 k_0)]^2$ smaller than the function plotted.
Except for a small region around $\theta=2$, where it diverges
(because $t_{min}^l \to \infty$), the function is always smaller than
its value for $\theta \to 0$\cite{note2}.
This means that correlations increase
the effect of the nonlinearity for almost all values of $\theta$.

\begin{section}{Solution in the noisy case}
So far we have neglected the presence of noise in Eq.~(\ref{eq1}).
We now turn to the study of the problem in presence both of deposition and
diffusion noise.
It will turn out that noise affects the position of the minimum only for
small initial roughness. For large and intermediate values of
$W_0$ the deterministic theory of Sec.~\ref{Det} is sufficient.

\label{Noise}
The inclusion of noise in the problem changes Eq.~(\ref{eqS}) to
\begin{equation}
\partial_t S(k,t) = -2 \left[\kappa k^4 - f[a(t)] k^2 \right] S(k,t) + 
R(k),
\label{eqSnoisy}
\end{equation}
with $R(k) = R_S + R_D k^2$.

The formal solution is
\begin{equation}
S(k,t) = S_{det}(k,t) + S_{noise}(k,t),
\end{equation}
where $S_{det}(k,t)$ is again given by Eq.~(\ref{formal}) while
\begin{equation}
S_{noise}(k,t) = R(k) S_{det}(k,t)
\int_0^t  ds \;\; S_{det}^{-1}(k,s).
\end{equation}

As before the key point is to realize that provided $k_m/k_0 \ll 1$,
after a short transient of duration $t_0 =1/(2 \kappa k_0^4)$ one can safely
take $S_{det}(k,t) \approx S_0 \exp \left(-2 \kappa k^4 t \right)$.
Using this expression
\begin{equation}
S_{noise}(k,t) = {R(k) \over 2 \kappa k^4} \left[1 -\exp \left(-2 \kappa k^4 t
 \right) \right].
\end{equation}
With this formula one can compute the additive contributions of the 
noisy part of the structure factor to $a(t)$ and to the roughness $W^2(t)$
which arise due to shot noise ($S$) and diffusion noise ($D$),
respectively, 
\begin{equation}
a(t) = a_{det}(t) + a_S(t) + a_D(t)
\end{equation}
and
\begin{equation}
W^2(t) = W^2_{det}(t) + W^2_S(t) + W^2_D(t).
\end{equation}

The results are
\begin{equation}
a_S(t) \approx {\pi R_S \over 4 \kappa} \log \left(2\kappa k_0^4 t \right)
\end{equation}
\begin{equation}
a_D(t) \approx {\pi R_D k_0^2 \over 2 \kappa}
\end{equation} 
and
\begin{equation}
W^2_S(t) \approx \pi R_S \sqrt{\pi t \over 2 \kappa}
\label{W2S}
\end{equation}
\begin{equation}
W^2_D(t) \approx {\pi R_D \over 4 \kappa} \log \left(2\kappa k_0^4 t \right).
\label{W2D}
\end{equation} 

The determination of the temporal evolution of the system is now more
complicated than in the noiseless case.
There, the form of $a(t)$ was always the same and the minimum for
$W^2$ changed depending on the various approximations for $f[a(t)]$.
Here, even the expression of $a(t)$ varies in time.
Moreover, in the noiseless case, the minimum in $W^2$ occurs when
the condition~(\ref{mintime}) is fulfilled.
Here, there are  three independent contributions to the width:
A minimum may appear much before the instability starts to play any role,
simply because of the interplay between the different contributions to $W^2$.
However, since $W_{det}^2$ is the only decreasing contribution, it is clear
that a minimum in $W^2$ can occur only if it already existed in the 
deterministic case. Noise cannot create a minimum and, as will be shown below
cannot destroy it, but only shift it to shorter times.
We will assume in the following that a minimum in the noiseless case
exists.

The complication related to the different contributions to $a(t)$
turns out to be unimportant for physical values of the parameters.
Inserting the expressions for $a_S$ and $a_D$ in Eq.~(\ref{current})
one sees that, under the physically sensible assumptions $l_D \gg a_\parallel$
and $k_0^{-1} \gg a_\parallel$, $f(a_S) \approx f(a_D) \approx \alpha$.
Hence, even if for long times it may happen that $a_S$ or $a_D$ become larger
than $a_{det}$, their value is so small that the theory is not affected.

The analysis of $W^2$ is instead quite involved.
Let us define the timescales
\begin{equation}
\tau_{RS} = {W_0^2 \over 2 k_0^2 \pi R_S} =
{Q^2 y^2 \over 32 \pi} \tau_l,
\end{equation}

\begin{equation}
\tau_{RD} = {2 \kappa W_0^4 \over \pi k_0^4 R_D^2} =
{Q^4 y^2 \over 8 \pi} \tau_l,
\end{equation}

\begin{equation}
\tau_{DS} = \left({R_D \over R_S}\right)^2 {1 \over 8 \pi \kappa} =
{y^2 \over 128 \pi} \tau_l.
\end{equation}

$\tau_{RS}$ is the timescale when $W_S$ becomes greater than $W_{det}$;
$\tau_{RD}$ is the timescale when $W_D > W_{det}$;
$\tau_{DS}$ is the timescale when $W_S > W_D$.

Let us carry out the analysis of the interplay of these different timescales
assuming, for the moment, that the deterministic part is well described by
linear theory, i. e. $t_{min}=t_{min}^l \approx 0.18 \tau_l$.

For small initial roughness ($Q \ll 1/2$) one finds that
$\tau_{RD} \ll \tau_{RS} \ll \tau_{DS} \ll \tau_l$ and this implies
that $W=W_{det}$ for $t \ll \tau_{RD}$,
$W=W_D$ for $\tau_{RD} \ll t \ll \tau_{DS}$,
$W=W_S$ for $\tau_{DS} \ll t \ll \tau_l$ and
$W \approx \exp(t)$ for $\tau_l \ll t$.
Hence the minimum occurs for $t_{min} \approx \tau_{RD}$ but before
the instability sets in there is another change for $t \approx t_{DS}$.

If instead $Q \gg 1/2$ there are two possibilities.
If the initial roughness is not very large ($Q^2 y^2 \ll 32 \pi$), one has
$\tau_{DS} \ll \tau_{RS} \ll \tau_l$ and
$W=W_{det}$ for $t \ll \tau_{RS}$,
$W=W_S$ for $\tau_{RS} \ll t \ll \tau_l$ and
$W \approx \exp(t)$ for larger times.
Otherwise, if $Q^2 y^2 \gg 32 \pi$, then
$\tau_{DS} \ll \tau_l \ll \tau_{RS} \ll \tau_{RD}$:
$W=W_{det}$ always, the noise is irrelevant and $t_{min}=t_{min}^l \approx
0.18 \tau_l$.
Notice that the condition $Q^2 y^2 \gg 32 \pi$ is, apart from the numerical
factor, the condition (15) of the paper by Krug and Rost~\cite{Krug99}.

We have so far assumed that $t_{min}^{det} = t_{min}^l$ of the same order
of magnitude of $\tau_l$.
When deterministic nonlinearities are strong they increase the value
of $t_{min}^{det}$, which becomes much greater than $\tau_l$.
As shown above, this starts to happen for $\tau_D \gg \tau_l$ which
corresponds to $Q^2 y^2 \gg (8 l_D/a_\parallel)^2 \gg 1$.
Then it might in principle happen that $t_{min}^{det}$ becomes larger
than $\tau_{RS}$ in the last of the cases above.
However, while $t_{min}^{det}$ grows for large $Q$ as $Q$ (Eq.~(\ref{tmin2}))
or as $Q^{4/3}$ (Eq.~(\ref{tmin3})), $\tau_{RS}$ is proportional to $Q^2$:
$\tau_{RS}$ always remains larger than $t_{min}^{det}$ and nothing in
the previous discussion changes.

In Table~\ref{Table1} is a summary of the different regimes found depending
on $Q$.

\end{section}

\section{Numerical results}
\label{Numerics}

The results presented above are obtained by analytically estimating
the behavior of the large-$N$ equation~(\ref{eqS}), which in turn
is an approximation of the fully nonlinear Eq.~(\ref{eqh}).
In order to check the validity of these results we have solved numerically 
the fully nonlinear equation for values of the parameters corresponding
to the different possible regimes of Table~\ref{Table1} and compared the
results with the numerical integration of the large-$N$ equation~\cite{note3}
and with the analytical estimates.
The numerical integration was performed by the simple first-order Euler
scheme, on a lattice of size $512 \times 512$. The temporal stepsize was
chosen to be 1, while the lattice spacing was equal to $\pi / k_0$,
with $k_0$ specified in the figure captions.

\begin{figure}
\centerline{\psfig{figure=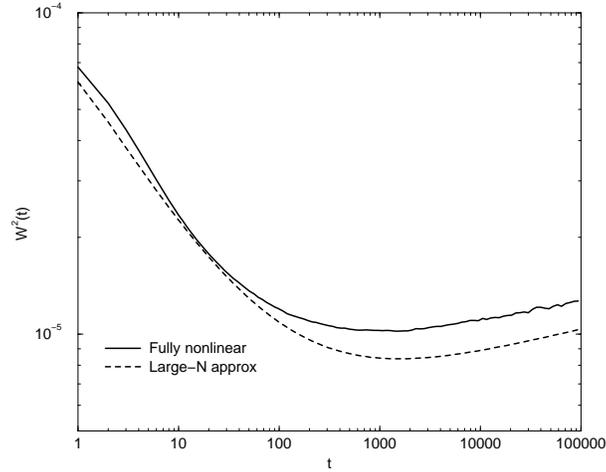,width=9cm,angle=-90}}
\caption{Double logarithmic plot of $W^2(t)$ vs $t$, for a system
with $k_0=1/10$, $W_0=1/100$, $Q=1/10$, $F=10^{-7}$,
$l_D=1000$ and $y=1/100$.
Here and in all other plots $a_\perp$ and $a_\parallel$ are taken to be
equal to 1.
For these values of the parameters $Q \ll 1/2$ and conserved 
noise dictates the position of the minimum.
}
\label{Fig2}
\end{figure}

We start by considering the limit of very small initial roughness, so that
$Q = 0.1$ and the minimum is due to conserved noise roughening the
surface (Fig.~\ref{Fig2}).
The minimum width for the fully nonlinear solution occurs for a time
compatible with the analytical prediction $t_{min}=\tau_{RD}\approx 1600$.
Despite having a minimum around the same time the solution of
the fully nonlinear case and the large-$N$ approximation differ
noticeably for large times. This poor agreement is however only
apparent and is due to a technical subtlety:
The numerical solution of the full equation is performed on a square lattice,
while the analytical calculations assume a circular Brillouin zone, $k < k_0$.
When noise is irrelevant, since the structure factor
decays exponentially for large wavevectors, the difference in
the Brillouin zones does not really matter after the initial transient $t_0$.
In the noise dominated cases instead, the structure factor has a
power-law tail: The effect of the different Brillouin zones persists in time,
cannot be eliminated easily~\cite{note} and leads to a systematic
overestimate of the value of $W^2(t)$.
This is why for long times the numerical solution is not in agreement
with the large-$N$ result. This problem is most evident in this case
dominated by conserved noise, as the power-law tail of the $S(k,t)$ is
broader. 

\begin{figure}
\centerline{\psfig{figure=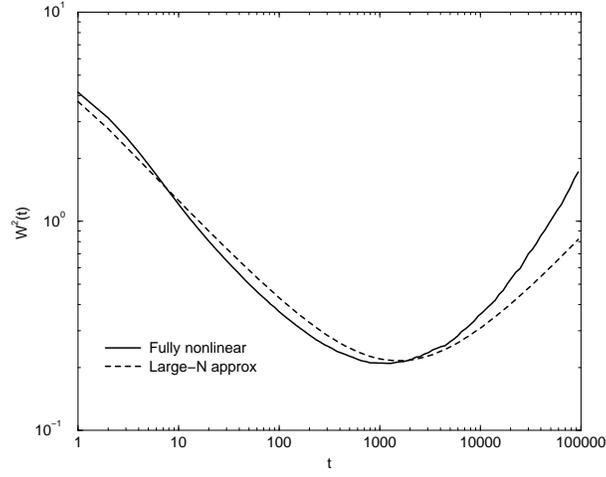,width=9cm,angle=-90}}
\caption{Double logarithmic plot of $W^2(t)$ vs $t$, for a system
with $k_0=1/40$, $W_0=10/4$, $Q=100$, $F=1$, $l_D=40$ and $y=1/100$.
For these values of the parameters $1/2 \ll Q \ll \sqrt{32 \pi}/y$:
The minimum is due to nonconserved noise.
}
\label{Fig3}
\end{figure}

In Fig.~\ref{Fig3} the value of the roughness of the fully nonlinear case and
of the large-$N$ approximation are plotted in the case where nonconserved
noise dominates.
Again, the analytical value $t_{min} = \tau_{RS} \approx 15700$ matches
quite well the numerical results.

\begin{figure}
\centerline{\psfig{figure=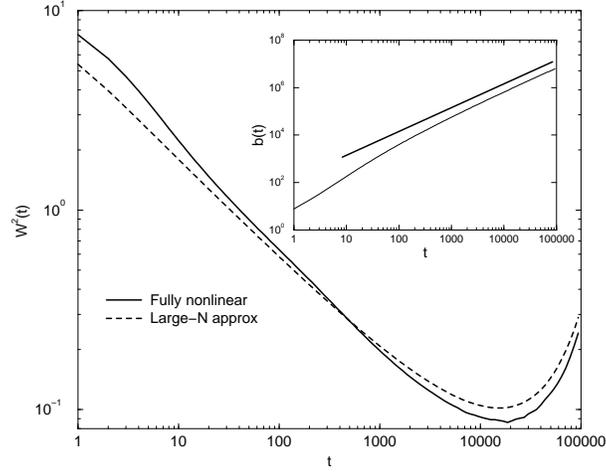,width=9cm,angle=-90}}
\caption{Double logarithmic plot of $W^2(t)$ vs $t$, for a system
with $k_0=1/100$, $W_0=3$, $Q=300$, $F=1/50$, $l_D=265.9$ and $y=1/10$.
The inset shows the log-log plot of $b(t)$ along with a line of slope 1.
For these values of the parameters $\sqrt{32 \pi}/y \ll Q \ll
8 l_D/(a_\parallel y)$: Linear noiseless theory holds.
}
\label{Fig4}
\end{figure}

The same quantities are reported in Fig.~\ref{Fig4} for values of
the parameters such that both noise and nonlinearities are irrelevant.
In this case, the analytical prediction for the position 
of the minimum is the one provided by linear theory $t_{min}^l
\approx 0.18 \tau_l \approx 14400$. Also in this case, the agreement between
numerics and the theoretical prediction is good. Notice that for these
values of the parameters the naive condition~(\ref{21Krug}) for the
irrelevance of nonlinearity is violated. However, the nonlinear contribution
to the current~(\ref{current}) is initially large but rapidly decays,
so that it does not influence the position of the minimum.
The initial deviation from linearity can also be seen in the behavior
of $b(t)$, plotted in the inset of Fig.~\ref{Fig4}.

\begin{figure}
\centerline{\psfig{figure=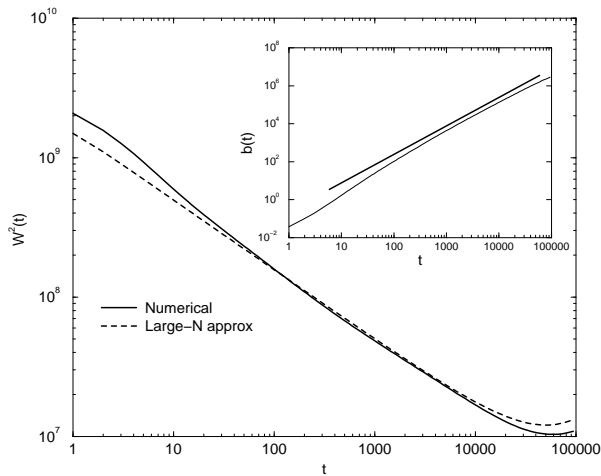,width=9cm,angle=-90}}
\caption{Double logarithmic plot of $W^2(t)$ vs $t$, for a system
with $k_0=1/100$, $W_0=50000$, $Q=5 \cdot 10^6$, $F=10^4$, $l_D=10$
and $y=1/1000$.
The inset shows the log-log plot of $b(t)$ along with a line of slope 3/2.
For these values of the parameters $8 l_D/(a_\parallel y) \ll Q \ll
8 l_D/(a_\parallel y^2)$ and the nonlinearity is relevant.
}
\label{Fig5}
\end{figure}

The position of the minimum is instead determined by the nonlinearity in
Fig.~\ref{Fig5}: Here linear theory would predict $t_{min}^l \approx 300$,
while the roughness keeps decreasing until $t \approx 60000$ in reasonable
agreement with the analytical estimate $t_{min} \approx 40000$ given by
Eq.~(\ref{tmin2}).
The nonlinear behavior is also evident in the inset, where $b(t)$ is
plotted: it is proportional to $t^{3/2}$ as predicted by Eq.~(\ref{bD}).

\begin{figure}
\centerline{\psfig{figure=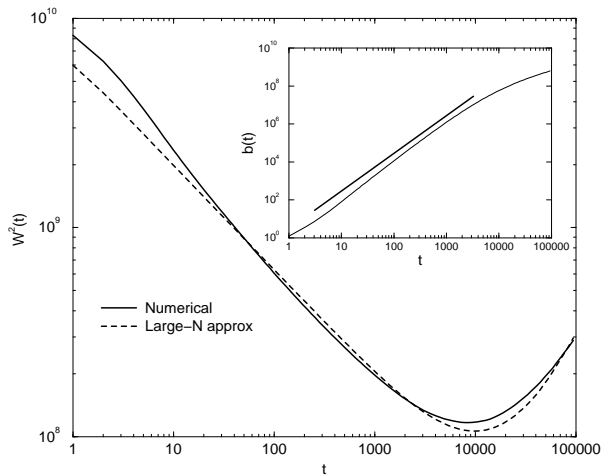,width=9cm,angle=-90}}
\caption{Double logarithmic plot of $W^2(t)$ vs $t$, for a system
with $k_0=1/1000$, $W_0=10^5$, $Q=10^8$, $F=10^4$, $l_D=100$
and $y=1/10$.
The inset shows the log-log plot of $b(t)$ along with a line of slope 2.
For these values of the parameters $8 l_D/(a_\parallel y^2) \ll Q$.
}
\label{Fig6}
\end{figure}

Finally, in Fig.~\ref{Fig6} the roughness in the most nonlinear case
($Q \gg 8 l_D/(a_\parallel y^2)$) is shown. In this case the linear
theory would predict $t_{min} \approx 0.03$, while Eq.~(\ref{tmin3})
yields $t_{min} \approx 5000$. The numerical result $t_{min} \approx 8000$
is again close to the estimate provided by nonlinear theory.
Consistently $b(t)$ grows as $t^2$ (Fig.~\ref{Fig6}, inset).

In summary, for initial values of the roughness ranging from
very small to very large we find that the evolution of the fully nonlinear
equation is well approximated by the large-$N$ limit and that the analytical
estimates found above agree with the numerical results.
The large-$N$ limit describes quite precisely this early
stage behavior because, up to the time when the minimum is reached,
the dynamics makes the surface smoother, reducing slope fluctuations
and making the approximation $(\nabla h)^2=\langle(\nabla h)^2\rangle$
increasingly more accurate.
Only after the minimum, when the instability takes over, slope fluctuations
grow, leading to the breakdown of the large-$N$ approximation.

\section{Discussion}
\label{Discussion}

In the previous Sections we have carried out a rather complete analysis
of the non monotonic behavior of the roughness of a surface which evolves
according to the continuum equation for unstable growth.
It must be stressed that an initial decrease of the
roughness is not necessarily due to a ``rough'' substrate, in the
sense of surface width exceeding some threshold. No matter
how small the substrate fluctuations, if they extend to a length
scale smaller than that of the linear instability, the roughness
will initially decrease. Only if substrate fluctuations are limited
to relatively large length scales, big amplitudes are needed.

The initial decrease of the roughness is
always governed by the relaxational Mullins-like term.
The moment when this initial decrease ends depends instead crucially
on the value of  $Q$ (i.e. on the initial roughness): the minimum may
be accounted for by linear (noisy or deterministic) or nonlinear theory.
Interestingly, for realistic values of the diffusion length $l_D$ and
of the ratio $y=l_{ES}/l_D$, the different regimes are nonoverlapping:
It is in principle possible to see each behavior by simply changing 
$W_0$, i. e. the initial roughness. 
It is clear, of course, that this
is not necessarily true in practice, since experimentally realizable
values of $Q$ are limited. Notice, however, that $Q$ depends on two 
quantities, $W_0$ and $k_0$, and also the variation of the latter
could help in expanding the range of variation of experimentally realizable
values of $Q$.
Moreover the presence of correlations in the initial roughness enhances
the effect of nonlinearity.

With regards to the experimental relevance of the present results
it is natural to compare them with the recent data of
Gyure {\em et al.}\cite{Gyure98}.
The comparison of linear theory with the same data~\cite{Krug99} pointed
out the violation of the naive condition for the irrelevance of
nonlinearity, Eq.~(\ref{21Krug}). As shown above the correct condition
for the irrelevance is different (Eq.~(\ref{21nonlin})) and turns out to
be fulfilled: The substrate is such that nonlinearity does not
matter. On the other hand, the experimental parameters give 
$Q \approx 110$ and $\sqrt{32 \pi}/y \approx 450$, indicating that
nonconserved noise mostly dictates the position of the minimum.
This conclusion is at odds with the results of Ref.~\cite{Krug99}.
The mismatch is due to a different treatment of numerical prefactors
and should
not be taken too seriously: The precision of values determined from
experimental data is poor and already introduces large uncertainty
in the physical parameters.
However, even if precise experimental data were available a very detailed
comparison between theory and experiment would not be possible, because
formulas linking parameters of the continuum equation ($\alpha, \kappa$)
with physical quantities ($l_D, l_{ES}$) are known only in order of magnitude.
Improved determination of the numerical prefactors would surely be
an important contribution to this field of research.

We have considered here only the form~(\ref{current}) of the unstable
current, which is valid in the limit of small ES barriers and does not
vanish for finite slopes. Other forms of the current are commonly used
for large ES barriers ($l_{ES} \gg l_D$) or when the current
vanishes for some ``magic'' value of the slope.
The previous analysis can be performed along the same lines for these
alternative currents.
We do not expect qualitatively different results. In particular, no
special behavior should be induced by the presence of magic slopes,
since the initial decay of fluctuations governed by the Mullins-like
term quickly washes out large slopes independently from the expression
of the current.

\vspace{0.5cm}

\noindent
{\bf Acknowledgements.} The support of the 
Alexander von Humboldt foundation (C.C.) and of DFG within SFB 237
(J.K.) is gratefully acknowledged. 


\begin{table}

\begin{center}
\begin{tabular}{||c|c|c||} \hline
Value of $Q$ & Position of the minimum & Relevant effect \\ \hline
$Q \ll 1/2$ & $\tau_{RD} = (Q^4 y^2/8 \pi) \tau_l$ & Conserved noise \\
$1/2 \ll Q \ll \sqrt{32 \pi}/y $ & $\tau_{RS} = 
(Q^2 y^2/32 \pi) \tau_l$ & Nonconserved noise \\
$\sqrt{32 \pi}/y \ll Q \ll 8 l_D/(a_\parallel y)$ & $0.18 \tau_l$ &
Linear deterministic \\
$8 l_D/(a_\parallel y) \ll Q \ll 8 l_D/(a_\parallel y^2)$ & $0.4
(\tau_l \tau_D)^{1/2}$ & Nonlinear \\
$8 l_D/(a_\parallel y^2) \ll Q $ & $0.5 (\tau_l \tau_D \tau_{ES})^{1/3}$
& Nonlinear \\
\hline
\end{tabular}
\end{center}
\caption{Analytical estimates of the position of the minimum depending
on the substrate roughness parameter $Q=W_0/(a_\perp a_\parallel k_0)$. 
The other quantities appearing in the Table
are determined by the characteristic length scales $l_D$ and $l_{ES}$
of the growing surface. In particular, $y = l_{ES}/l_D$ and 
$\tau_l = 4 \kappa/\alpha^2 \approx 16 F^{-1} (l_D/l_{ES})^2$.
For definitions of $\tau_D$ and $\tau_{ES}$ see Eqs.(\ref{tauD},
\ref{tauES}).}
\label{Table1}
\end{table}

\begin{references}
\bibitem{note2}
The limit of the function for $\theta \to 0$ is here $\pi/8$ and not 1
(as in Eq.~(\ref{21nonlin})) because $t_{min}^l$ is derived in a
different way and has a different numerical prefactor.
\bibitem{note3}
In the numerical integration of the large-$N$ equation, we solved
Eq.~(\ref{eqS}) with $a(t)$ given by the spatial average of $(\nabla h)^2$
resulting from the solution of the fully nonlinear equation.
\bibitem{note}
In principle one can compute $W^2(t)$ by integrating the spherically
averaged structure factor only for $k<k_0$. However, the aliasing phenomenon
implicit in taking the discrete Fourier transform~\cite{NumRec},
leads to a systematic overestimate of $S(k,t)$ for large wavevectors,
inducing a spuriously large roughness.
\end{references}
\end{document}